\documentclass[conference]{IEEEtran}
\IEEEoverridecommandlockouts
\usepackage{cite}
\usepackage{url}
\usepackage[super]{nth}
\makeatletter
\newcommand*{\rom}[1]{\expandafter\@slowromancap\romannumeral #1@}
\makeatother
\usepackage{amsmath,amssymb,amsfonts}
\usepackage{algorithmic}
\usepackage{graphicx}
\usepackage{textcomp}
\usepackage{xcolor}
\def\BibTeX{{\rm B\kern-.05em{\sc i\kern-.025em b}\kern-.08em
    T\kern-.1667em\lower.7ex\hbox{E}\kern-.125emX}}
\begin{document}

\title{Speech Recognition With No Speech Or With Noisy Speech Beyond English\\
{
}
\thanks{}
}

\author{\IEEEauthorblockN{Gautam Krishna}
\IEEEauthorblockA{\textit{Brain Machine Interface Lab} \\
\textit{The University of Texas at Austin}\\
Austin, Texas \\
}
\and
\IEEEauthorblockN{Co Tran}
\IEEEauthorblockA{\textit{Brain Machine Interface Lab} \\
\textit{The University of Texas at Austin}\\
Austin, Texas \\
}
\and
\IEEEauthorblockN{Yan Han}
\IEEEauthorblockA{\textit{Brain Machine Interface Lab} \\
\textit{The University of Texas at Austin}\\
Austin, Texas \\
}
\and
\IEEEauthorblockN{Mason Carnahan}
\IEEEauthorblockA{\textit{Brain Machine Interface Lab} \\
\textit{The University of Texas at Austin}\\
Austin, Texas \\
}
\and
\IEEEauthorblockN{Ahmed H Tewfik}
\IEEEauthorblockA{\textit{Brain Machine Interface Lab} \\
\textit{The University of Texas at Austin}\\
Austin, Texas  \\
}
}

\maketitle

\begin{abstract}
In this paper we demonstrate continuous noisy speech recognition using connectionist temporal classification (CTC) model on limited Chinese vocabulary using electroencephalography (EEG) features with no speech signal as input and we further demonstrate single CTC model based continuous noisy speech recognition on limited joint English and Chinese vocabulary using EEG features with no speech signal as input. We demonstrate our results using various EEG feature sets recently introduced in \cite{krishna2019state} as well as we propose a deep learning architecture in this paper which can perform continuous speech recognition using raw EEG signals on limited joint English and Chinese vocabulary. 
\end{abstract}


\section{Introduction}

Electroencephalography (EEG) is a non invasive way of measuring electrical activity of human brain. In \cite{krishna2019speech} authors demonstrated deep learning based automatic speech recognition (ASR) using EEG signals for a limited English vocabulary of four words and five vowels. In \cite{krishna20} authors demonstrated continuous noisy speech recognition using EEG for larger English vocabulary using connectionist temporal classification (CTC) model and attention model \cite{chorowski2015attention}. We use only CTC model in this work. 
In this paper we extend the work reported in \cite{krishna2019state,krishna20} for a much larger Chinese vocabulary and joint Chinese English or multilingual vocabulary.
In \cite{krishna2019speech,krishna2019state,krishna20} authors demonstrated results using EEG features but in this work we also introduce a new deep learning architecture which can perform continuous speech recognition using raw EEG signals. 
Inspired from the unique robustness to environmental artifacts exhibited by the human auditory cortex \cite{yang1991auditory,mesgarani2011speech} we used very noisy speech data for this work and demonstrated lower character error rate (CER) for smaller corpus size using EEG features.

In \cite{wang2017simulation} authors decode imagined speech from EEG using synthetic EEG data and CTC network but in our work we use real EEG data, use multilingual vocabulary. In \cite{kumar2018envisioned} authors perform envisioned speech recognition using random forest classifier but in our case we use end-to-end state of art model and perform recognition for noisy speech. In \cite{ramsey2017decoding} authors demonstrate speech recognition using electrocorticography (ECoG) signals, which are invasive in nature but in our work we use non invasive EEG signals.

References \cite{gonzalez2015real,waibel2000multilingual,toshniwal2018multilingual} indicates some of the prior work done in the field of multilingual speech recognition but none of the prior work used EEG signals for performing recognition. In \cite{toshniwal2018multilingual} authors use a single end-to-end attention model for performing recognition but in our work used a single CTC model for performing multilingual speech recognition.

References \cite{amodei2016deep,yu2003chinese} explains some of the prior work done on Chinese and English joint speech recognition but EEG features were not used for performing recognition. 

One of the unique ability of human brain is multilingualism \cite{costa2014does,higby2013multilingualism}, ie: our brain is capable of understanding and speaking out multiple languages. This was another motivating factor for this work. All the subjects who took part in the experiments were multilingual.

We believe speech recognition using EEG will help people with speaking difficulties to use voice activated technologies with better user experience. As demonstrated in \cite{krishna2019speech} EEG helps ASR systems to overcome performance loss in presence of background noise. This will help ASR systems to perform with high accuracy in very noisy environments like airport, shopping mall etc where there is high level of background noise. Developing a robust multilingual speech recognition system using EEG will help in improving technology accessibility for multilingual people with speaking disabilities. In this work we use Chinese and English languages, which have zero overlap in their scripts and very noisy data was used. Hence we investigate one of the most challenging cases of multilingual speech recognition in this paper.

Major contribution of this paper is the extension of results presented in \cite{krishna2019state,krishna20} for a larger Chinese corpus, joint English Chinese corpus using different EEG feature sets as well as demonstration of multilingual speech recognition using raw EEG.

\section{Connectionist Temporal Classification (CTC)}

We used a single layer of gated recurrent unit (GRU) \cite{chung2014empirical} with 128 hidden units as the encoder of our CTC ASR model. The decoder of the CTC model consists of a dense or fully connected layer at every time step and a softmax activation function. The output of the GRU encoder is fed into the decoder part of the network at every time step. The dense layer performs affine transformation. The softmax activation function outputs character prediction probabilities at every time step. 

The encoder GRU takes EEG features as input and transforms it into hidden representations. The decoder part of the network takes these hidden representations as input and transforms it into output text.
We used a character based model for this work. The model was predicting a character at every time step. The model was trained for 400 epochs using adam \cite{kingma2014adam} optimizer with a batch size one to optimize the CTC loss. The mathematical details of the CTC loss function is covered in \cite{graves2014towards,graves2006connectionist,krishna20,krishna2019state}.  A dynamic algorithm is used to compute the CTC loss. 

During inference time we used CTC beam search decoder. The architecture of our CTC model is shown in Figure 1. The Figure 3 shows the CTC model training loss convergence.


\begin{figure}[h]
\begin{center}
\includegraphics[height=8.5cm, width=\linewidth,trim={0.1cm 0.1cm 0.1cm 0.1cm}]{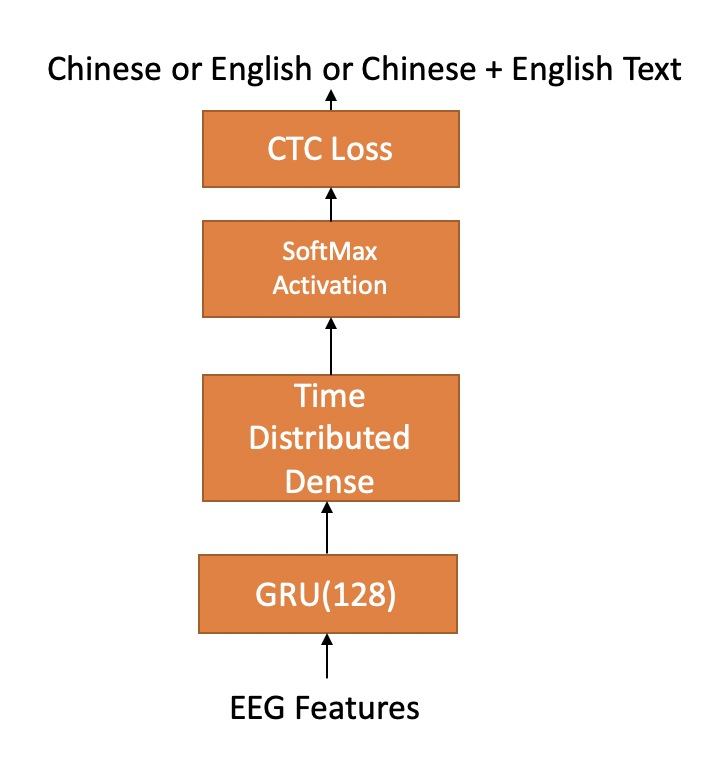}
\caption{CTC Model for EEG feature set input} 
\label{1vsall}
\end{center}
\end{figure}

\section{Speech Recognition using Raw EEG}

The CTC model explained in previous section was used when the input to the model was EEG features but not for raw EEG. Raw EEG signals refer to the EEG signals just after preprocessing (removing artifacts) but before extracting features and dimension reduction. 
For training the CTC model with raw EEG signal we modified the encoder network architecture in the CTC model. 
We added a combination of one dimensional convolutional layer followed by one dimensional max pooling layer followed by another one dimensional convolutional layer followed by another one dimensional max pooling layer at the input and this network was connected to the GRU layer in the default CTC network encoder.
No modification was made to decoder architecture. The motivation behind using convolutional layers was to see if the model can discover better representations from the raw EEG signals rather than feeding the model with hand craft EEG features. We used a development set to figure out the hyper parameters for the convolutional and max pooling layer. 

For both one dimensional convolutional layers we used 100 filters, kernel size of 3, padding type 'same' and relu activation. For both one dimensional max pooling layers we used pool size of 2, stride of one and padding type 'same'. The model architecture is described in Figure 4.

\section{Design of Experiments for building the database}

We built simultaneous speech EEG recording English and Chinese databases for this work. Five female and seven male subjects took part in the experiment. All subjects were UT Austin undergraduate,graduate students in their early twenties. All subjects were native Mandarin Chinese speakers and English was their foreign language.

 The 12 subjects were asked to speak 10 English sentences and their simultaneous speech and EEG signals were recorded. The first 9 English sentences were the first 9 sentences from the USC-TIMIT database\cite{narayanan2014real}, while the \nth{10} sentence was " Can I get some water ".
 
 This data was recorded in presence of background noise of 65 dB. Background music played from our lab desktop computer was used as the source of noise. We then asked each subject to repeat the same experiment two more times, thus we had 36 speech EEG recording examples for each sentence.
 
We then asked the 12 subjects to repeat the same set of previous experiment but this time they were asked to speak the Chinese translation of the 10 English sentences.

We used Brain Vision EEG recording hardware. Our EEG cap had 32 wet EEG electrodes including one electrode as ground as shown in Figure 2. We used EEGLab \cite{delorme2004eeglab} to obtain the EEG sensor location mapping. It is based on standard 10-20 EEG sensor placement method for 32 electrodes.

For this work, we used data from first 10 subjects for training the model, remaining two subjects data for validation and test set respectively.

\begin{figure}[h]
\begin{center}
\includegraphics[height=3cm,width=0.25\textwidth,trim={1cm 1cm 1cm 0.1cm},clip]{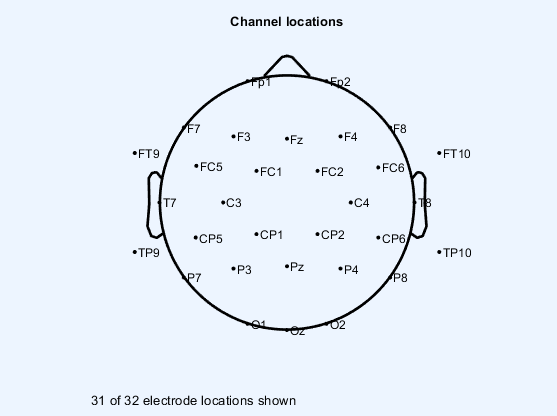}
\caption{EEG channel locations for the cap used in our experiments} 
\label{1vsall}
\end{center}
\end{figure}

\begin{figure}[h]
\begin{center}
\includegraphics[height=5cm,width=0.3\textwidth,trim={0.1cm 0.1cm 0.1cm 0.1cm},clip]{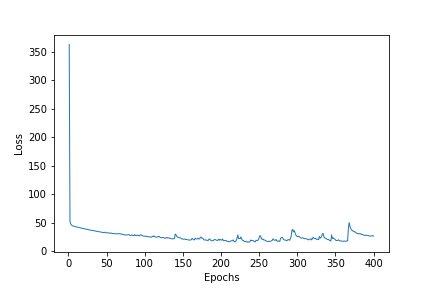}
\caption{CTC training loss convergence when trained with EEG feature set} 
\label{1vsall}
\end{center}
\end{figure}


\begin{figure}[h]
\begin{center}
\includegraphics[height=8.5cm, width=\linewidth,trim={0.1cm 0.1cm 0.1cm 0.1cm}]{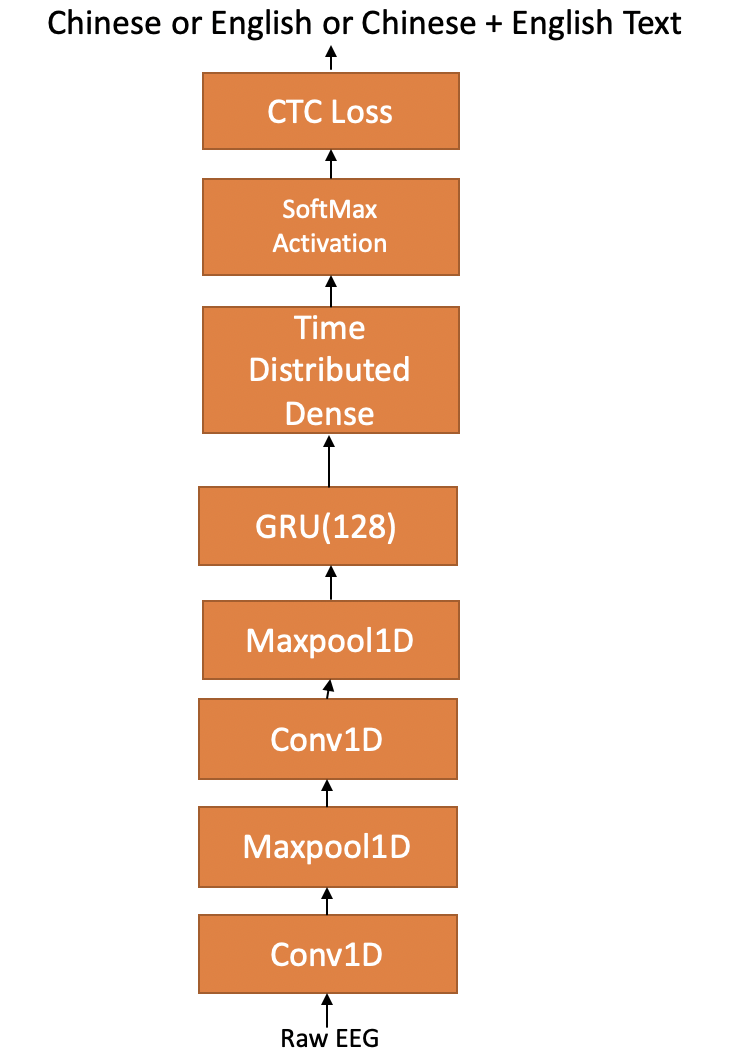}
\caption{CTC Model for Raw EEG input} 
\label{1vsall}
\end{center}
\end{figure}

\section{EEG and Speech feature extraction details}

We followed the same preprocessing methods used by authors in \cite{krishna20,krishna2019speech,krishna2019state} to preprocess the EEG, speech signals and for extracting EEG and acoustic features. 

EEG signals were sampled at 1000Hz and a fourth order IIR band pass filter with cut off frequencies 0.1Hz and 70Hz was applied. A notch filter with cut off frequency 60 Hz was used to remove the power line noise.
EEGlab's \cite{delorme2004eeglab} Independent component analysis (ICA) toolbox was used to remove other biological signal artifacts like electrocardiography (ECG), electromyography (EMG), electrooculography (EOG) etc from the EEG signals.

We extracted three EEG feature sets described in \cite{krishna2019state} after preprocessing the EEG signals. All EEG features were extracted at a sampling frequency of 100Hz. The dimension of EEG feature set 1 was 155, feature set 2 and 3 had dimension 93. Details of EEG feature sets are covered in \cite{krishna2019state}. 

The recorded speech signal was sampled at 16KHz frequency. We extracted Mel-frequency cepstrum coefficients (MFCC) as features for speech signal.
We first extracted MFCC 13 features and then computed first and second order differentials (delta and delta-delta) thus having total MFCC 39 features.
The MFCC features were also sampled at 100Hz same as the sampling frequency of EEG features to avoid seq2seq problem.

\section{EEG Feature Dimension Reduction Algorithm Details}

After extracting EEG and acoustic features as explained in the previous section, we used non linear methods to do feature dimension reduction in order to obtain set of EEG features which are better representation of acoustic features as explained by authors in \cite{krishna2019speech}. The tool we used for this purpose was Kernel Principle Component Analysis (KPCA) \cite{mika1999kernel}. We plotted cumulative explained variance versus number of components to identify the right feature dimension for each feature set. We used KPCA with polynomial kernel of degree 3 \cite{krishna2019speech,krishna20,krishna2019state}. The final dimensions of feature set 1, 2 and 3 after dimension reduction and computing delta, delta-delta were 90, 150 and 279 respectively. More details are covered in \cite{krishna2019state}.

\section{Results}

We used character error rate (CER) as performance metric to evaluate the our CTC model since it was predicting a character at every time step. The CTC model was trained for 400 epochs to observe loss convergence and batch size was set to one. Table 2 shows the results for recognition of Chinese sentences during test time using EEG feature set 1 for GRU encoder with 128 hidden units and 64 hidden units . As seen from the table, the error rate goes up as the vocabulary size increases. When the model was trained on first 3 sentences from training set and tested on first 3 sentences from the test set with GRU 128 units encoder, a low CER of \textbf{1.38 \%} was observed. For number of sentences = \{7,10\}, we also tried training the model with concatenation of \textbf{MFCC} and \textbf{EEG} features and observed error rates \textbf{55.36 \%} and \textbf{66.11 \%} on test set respectively, which were slightly lower than error rates observed when the model was trained using only EEG features as seen from Table 2. In general we observed that as the vocabulary size increases, adding MFCC features to EEG features will help in reducing the error rates.

For the multilingual training scenario, we are given two languages English and Chinese, each with independent character sets \{$C_1$, $C_2$\} and training sets \{($X_1$,$Y_1$), ($X_2$,$Y_2$)\}. Then the combined training data set is given by the union of the data sets for each language, $\bigcup\limits_{i=1}^{2} (X_{i},Y_{i})$ and the character set for the combined data set is given by $\bigcup\limits_{i=1}^{2} C_{i}$. Table 1 shows the result obtained for multilingual speech recognition using  EEG feature set 1,2,3 and raw EEG with GRU 128 hidden units encoder RNN. Again, lower error rates were observed for smaller corpus sizes and error rate went up as we increase the corpus size to 7 or 10 sentences. 
We believe as Chinese vocabulary has large number of unique characters, the model needs more number of training examples to generalize better and to give lower CER as corpus size increases.

In \cite{krishna2019speech} authors demonstrated that EEG sensors T7 and T8 contributed most to ASR test time accuracy, so we tried training the model with EEG feature set 1 from only T7 and T8 sensors for Chinese corpus and we observed that for some examples of corpus size, the error rates were comparable withe error rates shown in Table 2. We observed error rates \{58.7 \%,69.7 \%,70.8\%,70.1\% \} for number of sentences=\{3,5,7,10\} respectively with GRU 128 hidden units encoder. 

We tried reducing the number of hidden units of the GRU to 64 to see if it can help with overcoming the performance loss due to less amount of training examples and obtained results are shown in Table 2. The results indicate lower CER for larger corpus size compared to the results obtained using GRU encoder with 128 hidden units. Similarly for T7, T8 training with GRU 64 hidden units on Chinese vocabulary, we observed little reduced CER values of \textbf{52.5 \%} and \textbf{69.5 \%} for number of sentences=\{3,7\} respectively. For number of sentences =\{5,10\} the error rates were nearly same as the rates obtained using GRU 128 hidden units encoder. 

When we trained GRU 64 hidden units model on multilingual vocabulary with EEG feature set 1 we observed a lower error rate of \textbf{44.4 \%} for number of sentences = \{3\}, for number of sentences = \{5,7,10\} the error rates were nearly same as the rates reported in Table 1. 

We further observed that for the joint Chinese English or for the multilingual vocabulary, 2 layer GRU with 64 hidden units in each layer model gave lower error rates of \textbf{61.3 \%,72.4\%,74.3\%} for number of sentences equal to \{5,7,10\} respectively with EEG feature set 1 compared to the error rates reported in Table 1. 

\begin{table}[!ht]
\centering
\begin{tabular}{|l|l|l|l|l|l|}
\hline
\textbf{\begin{tabular}[c]{@{}l@{}}No \\ of\\ Sen\end{tabular}} & \textbf{\begin{tabular}[c]{@{}l@{}}No\\  of\\ unique\\ charac\\ ters\end{tabular}} & \textbf{\begin{tabular}[c]{@{}l@{}}Feature\\ Set 1\\ EEG\\ CER\%\end{tabular}} & \textbf{\begin{tabular}[c]{@{}l@{}}Feature\\ Set 2\\ EEG\\ CER\%\end{tabular}} & \textbf{\begin{tabular}[c]{@{}l@{}}Feature\\ Set 3\\ EEG\\ CER\%\end{tabular}} & \textbf{\begin{tabular}[c]{@{}l@{}}Raw\\ EEG\\ CER\%\end{tabular}} \\ \hline
3                                                               & 43                                                                                 & \textbf{45.6}                                                                  & 58                                                                             & 50                                                                             & 69                                                                 \\ \hline
5                                                               & 57                                                                                 & 66.8                                                                           & \textbf{56.2}                                                                  & 57.8                                                                           & 58.6                                                               \\ \hline
7                                                               & 81                                                                                 & 78.4                                                                           & 79.8                                                                           & \textbf{75.5}                                                                  & 78.2                                                               \\ \hline
10                                                              & 111                                                                                & 79.4                                                                           & 72.9                                                                           & \textbf{72.2}                                                                  & 75.5                                                               \\ \hline
\end{tabular}
\caption{CER on test set for \textbf{CTC model} for joint \textbf{Chinese English vocabulary}}
\end{table}

\begin{table}[!ht]
\centering
\begin{tabular}{|l|l|l|l|}
\hline
\textbf{\begin{tabular}[c]{@{}l@{}}Number of\\ Sentences\end{tabular}} & \textbf{\begin{tabular}[c]{@{}l@{}}Number of\\ unique\\ characters\\ contained\end{tabular}} & \textbf{\begin{tabular}[c]{@{}l@{}}GRU(128)\\ EEG\\ (CER\%)\end{tabular}} & \textbf{\begin{tabular}[c]{@{}l@{}}GRU(64)\\ EEG\\ (CER\%)\end{tabular}} \\ \hline
3                                                                      & 24                                                                                           & \textbf{1.38}                                                             & 3.6                                                                      \\ \hline
5                                                                      & 37                                                                                           & 34.7                                                                      & \textbf{31.6}                                                            \\ \hline
7                                                                      & 59                                                                                           & 64.8                                                                      & \textbf{49.6}                                                            \\ \hline
10                                                                     & 88                                                                                           & 69.4                                                                      & \textbf{65.8}                                                            \\ \hline
\end{tabular}
\caption{CER on test set for \textbf{CTC model} for \textbf{Chinese vocabulary} using \textbf{feature set 1}}
\end{table}


\section{Conclusions}

In this paper we demonstrated the feasibility of using EEG feature sets for performing continuous speech recognition to recognize Chinese and joint Chinese English or multilingual vocabulary. 
To the best of our knowledge this is the first time a continuous speech recognition using only EEG features is demonstrated for Chinese and multilingual vocabulary. Our results demonstrate that multilingual speech recognition with GRU 128 hidden units RNN encoder using EEG feature set 3 \cite{krishna2019state} gave the lowest test time error rate for maximum corpus size.  As corpus size increase speech recognition performance using raw EEG was comparable with results obtained using EEG features.  

We observed that as corpus size increase, CTC model CER went up and concatenating acoustic features with EEG features will help in reducing CER. Our work demonstrates the feasibility of using EEG features for performing multilingual speech recognition.

We further plan to publish our speech EEG data base used in this work to help advancement of the research in this area. For future work, we plan to build a much larger speech EEG data base and investigate whether CTC model results can be improved by training with more number of examples, by incorporating an external language model during inference time or by including a language identification model.

\section{Acknowledgements}

We would like to thank Kerry Loader and Rezwanul Kabir from Dell, Austin, TX for donating us the GPU to train the CTC model used in this work.

\bibliographystyle{IEEEtran}

\bibliography{refs}
\end{document}